\documentclass[%
 reprint,
 amsmath,amssymb,
 aps,
 aapm,
]{revtex4-2}

\usepackage{graphicx}
\usepackage{dcolumn}
\usepackage{bm}
\usepackage[dvipsnames]{xcolor}
\usepackage{subfig}
\usepackage{ulem}

\usepackage{caption}
\begin{document}

\preprint{APS/123-QED}

\title{A self-sustaining mechanism for Internal Transport Barrier formation in HL-2A tokamak plasmas}

\author{W. H. Lin$^{1}$, J. Garcia$^{2}$, J. Q. Li$^{1,b}$, S. Mazzi$^{2}$, Z. J. Li$^{1}$, X. X. He$^{1}$, X. Yu$^{1}$}
\affiliation{
${^1}$Southwestern Institute of Physics, Chengdu 610041, China.
\\${^2}$CEA, IRFM, F-13108 Saint Paul-lez-Durance, France.
\\$^b$Author to whom correspondence should be addressed: lijq@swip.ac.cn
}%

\date{\today}

\begin{abstract}

The formation of Internal Transport Barrier (ITB) is studied in HL-2A plasmas by means of nonlinear gyrokinetic simulations. A new paradigm for the ITB formation is proposed in which different physics mechanisms play a different role depending on the ITB formation stage. In the early stage, fast ions, introduced by Neutral Beam Injection (NBI) ion system, are found to stabilize the thermal-ion-driven instability by dilution, thus reducing the ion heat fluxes and finally triggering the ITB. Such dilution effects, however, play a minor role after the ITB is triggered as electromagnetic effects are dominant in the presence of established high pressure gradients. We define the concept of ITB self-sustainment, as the low turbulence levels found within the fully formed ITB are consequences of large scale zonal flows, which in turn are fed by a non-linear interplay with large scale high frequency electromagnetic perturbations destabilized by the ITB itself.

\end{abstract}

\maketitle


\section{Introduction}
The final goal of magnetic confinement devices is to confine plasmas of high temperature and density for sufficiently long time in order to produce economically advantageous fusion energy. Confined plasmas can be severely degraded by the outward energy transport driven by micro-instabilities such as the Ion-Temperature-Gradient (ITG) mode \cite{romanelli1989ion}. Therefore, a credible path towards reliable energy fusion production must rely on mechanisms controlling such an energy transport. \par
Plasmas with Internal transport barriers (ITB) \cite{connor2004itb}, characterized by a suppression of heat transport driven by microturbulence leading to high core temperatures and densities, have been shown to provide a way to improve plasmas energy confinement in various tokamaks \cite{bell1998core, romanelli2010fast, doyle2002progress, joffrin2003internal, sakamoto2004properies, yu2016}. The formation and characteristics of ITB have been extensively studied. Several physical mechanisms have been put forward to explain energy transport reduction or suppression within an ITB. One of the initial mechanisms proposed was the $E\times B$ flow shear turbulence stabilization (see \cite{connor2004itb} for example), which manifests itself by breaking up turbulent eddies and reducing the amplitude and cross phase of turbulent fluctuations. In this context, negative or low magnetic shear is also known to have a synergistic effect with $E\times B$ shear on ITB formation, as it weakens the drive of some unfavourable instabilities \cite{diamond1997dynamics} on one hand and prevents the detrimental effects brought by $E\times B$ shear \cite{burrell1997effects} on the other. Other mechanisms related to the presence of highly energetic fast ions have been proposed as well. A large fraction of fast ions produced from neutral beam injection (NBI) are found to be crucial in ITB formation by their dilution effects \cite{tardini2007thermal}, while a small minority of them could also be decisive through mechanisms such as linear resonant interaction with ITG \cite{siena2021new} or the enhancement of $\alpha$-stabilization \cite{romanelli2010fast}.\par
Despite the amount of studies devoted to clarify the physical mechanism behind the ITB formation, there are still aspects that remain unclear, e.g., whether a single physical mechanism or multiple ones are responsible for the ITB triggering and whether such mechanisms play significant roles on the ITB sustainment once it is fully formed. Clarifying these aspects is essential in order to properly evaluate whether plasmas with ITBs will be possible in future fusion reactors, for which some mechanisms, such as the $E\times B$ shearing produced by external injected torque, are known that will be less efficient.\par
\begin{figure*}
	\includegraphics[width=0.7\textwidth]{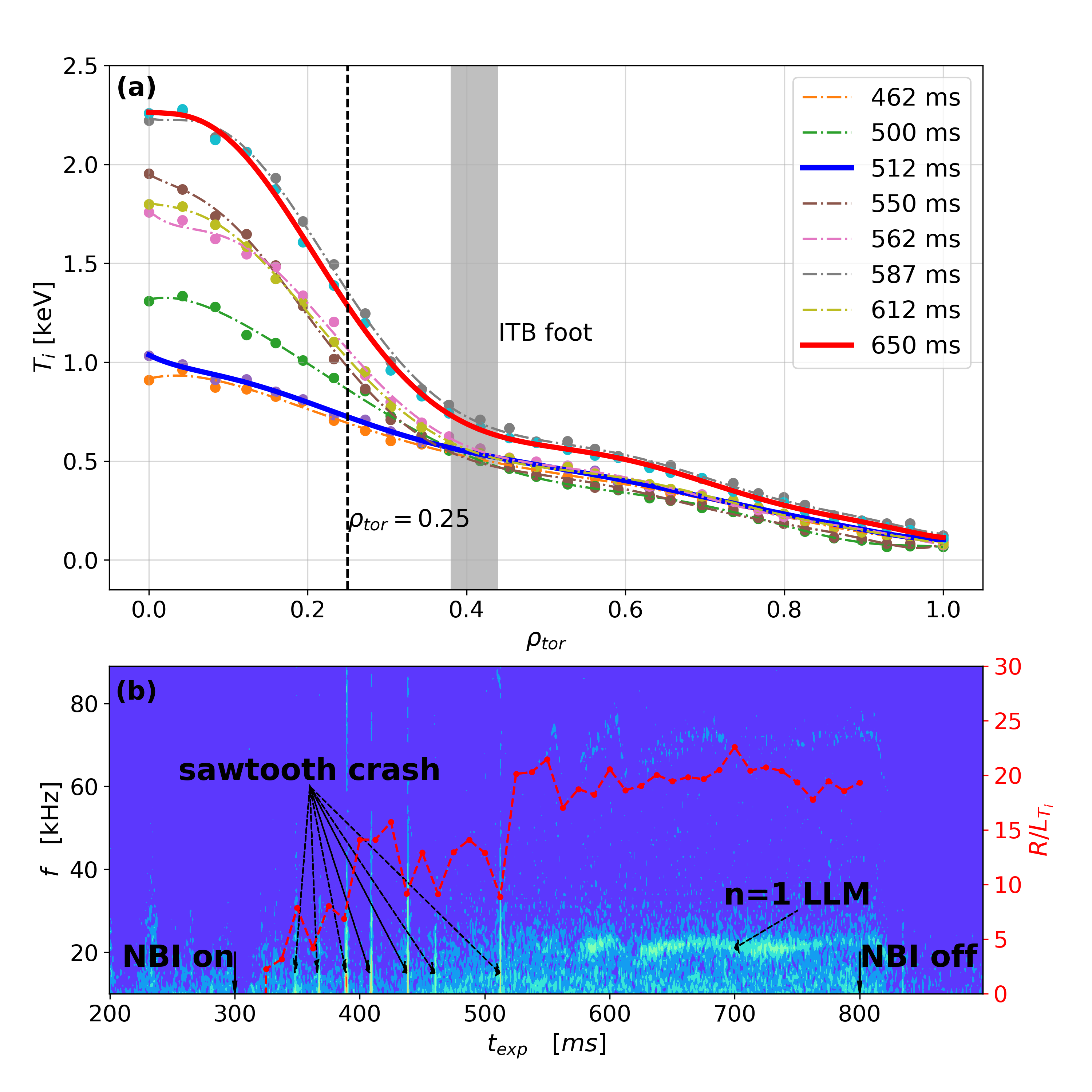}
	\caption{(a) Profiles of ion temperature $T_i$ measured by CXRS \cite{wei2014high} at different time points in HL-2A discharge \#22453. (b) The frequency spectrum of mirnov coil signal and the normalized logarithmic ion temperature gradient $R/L_{T_i}$ at $\rho_{tor}=0.25$ in the same discharge.}
	\label{LLM_in_experiment}
\end{figure*}
In this work, it is shown that the triggering and sustainment of ITB rely on two different physical mechanisms depending on the ITB formation stage. Whereas the ITB triggering is found to be a consequence of the NBI fast-ion dilution, it is proposed the concept of self-sustainment of the ITB as it is the ITB itself producing the physical mechanisms that provides its sustainment. The increase of electromagnetic (EM) effects in the presence of strong ITB-generated pressure gradients reduces turbulence and transport through the onset of large scale zonal flows \cite{diamond2005zonal} (with toroidal number $n=0$ and frequency $\omega=0$), which tap energy non-linearly from large scale MagnetoHydroDynamics (MHD) fluctuations that are destabilized by the ITB itself. Meanwhile, the $E\times B$ shearing generated by the plasma rotation is not found to play a major role on the ITB formation. Such findings may pave the way for the formation of ITBs in future tokamaks as long as EM effects are dominant. \par
Turbulence and transport analyses are performed with state-of-the-art gyrokinetic simulations for the ITB discharge \#22453 in the HL-2A tokamak {\cite{yu2016}}. In such a discharge as shown in Fig. \ref{LLM_in_experiment}, the ITB is triggered at about t=510ms, and stably sustained for a time window of about $250$ ms. During the ITB formation, the core ion temperature has increased from 1.0 to 2.3 $keV$, forming a region of large $R/L_{T_i}$ ($\approx20$) with the ITB foot located at $\rho_{tor} \approx 0.4$. Here, R is the major radius, $L_{T_i}$ the inverse logarithmic gradient of ion temperature and $\rho_{tor}$ the normalized square root of toroidal magnetic flux. The profiles at 510 and 650 ms, when the ITB begins to trigger and has been fully developed, respectively, are of particular interest to our study and provide the parameters set for the simulations discussed below. Shortly after the ITB triggering, the Mirnov coils detect two perturbations, a weak one at 70 kHz and a stronger one at 20 kHz in the laboratory frame, the latter being identified as a long-lived mode (LLM) in previous work \cite{Zhang2014}. LLMs, as well as fishbone (FB) instabilities \cite{yu2017transition}, are both MHD modes frequently observed in HL-2A after ITB triggering. Although FBs are proposed as the key factors of the ITB formation in some tokamaks \cite{günter2001mhd, ge2023multiple, wang2023theinternal}, they could hardly be related to the ITB triggering in HL-2A \cite{He2022} where FBs are less observed preceding the ITB. As for LLMs, a limited amount of works \cite{deng2022investigation} exist regarding their effects on ITB. The dynamic interplay between LLM and ITB remains obscure so far and will be investigated further in this work. \par
The structure of this paper is arranged as follow: after the simulation setup is addressed in section 2, the dominant instabilities in various simulation conditions are analyzed in section 3, and the stabilizing factor that is of vital importance on ITB formation is investigated by the analysis of ion heat flux in section 4. It will be shown that the full ITB formation benefitted from not only the linear stabilization of dominant instabilities but also the nonlinear EM effect, which is attributed in section 5 to the onset of zonal flow through the saturation of large scale EM modes such as the aforementioned LLM. Finally in section 6, the mechanisms governing the ITB formation on different stages are concluded and a full picture of ITB’s self-sustainment is proposed.\par

\section{Simulation Setup}
All simulations reported in this paper are performed with the first-principle gyrokinetic code GENE \cite{jenko2000electron} in flux-tube version. The simulated flux-tube is at $\rho_{tor,0}=0.25$, slightly inside the ITB foot. Here the subscript ‘0’ indicates flux-tube location. Miller geometry \cite{miller1998noncircular} is extracted from EFIT equilibrium. An extended region of low but positive magnetic shear $\hat{s}$ is observed inside the ITB foot, and at the simulated location $\hat{s}=0.12$ with the safety factor $q_0=1.05$. Typical grid parameters are as follows: perpendicular box sizes $[L_x,L_y]=[272,218]$ in units of ion Larmor radius $\rho_i$ with discretizations $[n_x,n_y]=[768,48]$, $n_z=32$ points in parallel direction, 32 points in parallel velocity directions and 32 magnetic moments. Here, x is the radial coordinate defined as $x=a\rho_{tor}$ ($a$ the minor radius), y the binormal coordinate and z the coordinate along the field line. When the effects of the perpendicular flow shear are considered, large aspect ratio and circular poloidal cross-section are assumed and therefore the normalized mean $E\times B$ shearing rate is defined as $\gamma_{E}\equiv(\rho_{tor,0}/q_0)(d\Omega/d\rho_{tor})/(c_s/R)$. $\Omega$ is the toroidal angular velocity, R the major radius and $c_s$ the sound speed. The full impact of $\gamma_{E}$ is considered in non-linear simulations only, in order to ensure the compatibility of the $E \times B$ algorithm \cite{McMillan2019Simulating} implemented in GENE. Other physical parameters are shown in Table \ref{tab:parameters}. With the aim of analyzing the individual effects of fast ion, finite-$\beta$ and $\gamma_E$, simulations are divided into subsets with or without some of these parameters, and they are performed at both the ITB triggering time, 510 ms, and when the ITB is well-developed at 650 ms. Note that the tuple $(n_{fi}, \beta, \gamma_{E} )$ is frequently used in the following figures to indicate the simulation conditions.\par
\begin{figure*}
	\includegraphics[width=0.95\textwidth]{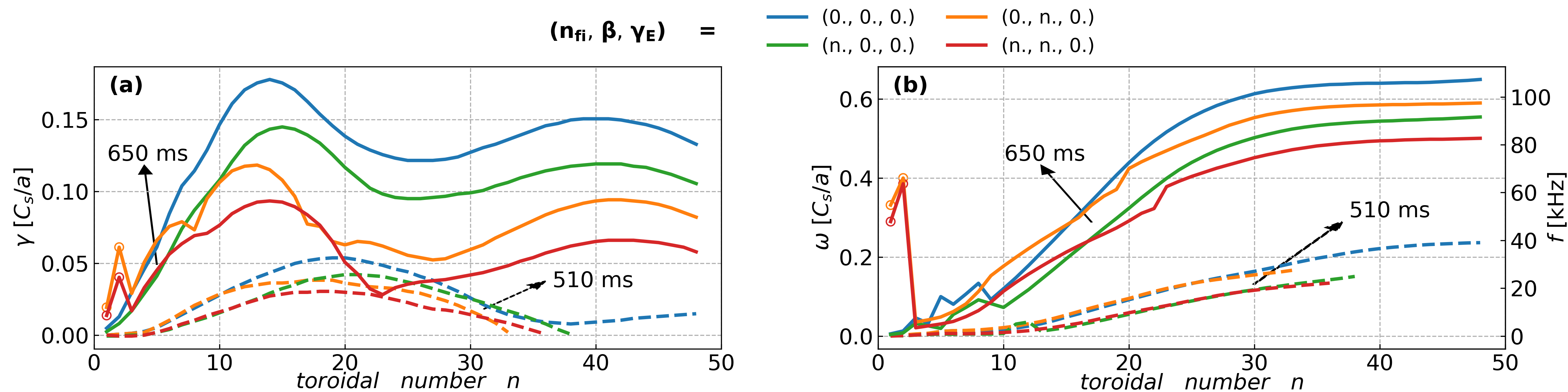}
	\caption{Spectrum of growth rates (a) and frequencies (b) in linear simulations for the cases with $(n_{fi}, \beta, \gamma_{E} )$ labeled above; here, 'n.' indicates the quantity is set to its nominal values in Table \ref{tab:parameters} and such notation will be used in the following figures.}
	\label{linear_instabilities_new}
\end{figure*}
\begin{figure*}
    \includegraphics[width=0.95\textwidth]{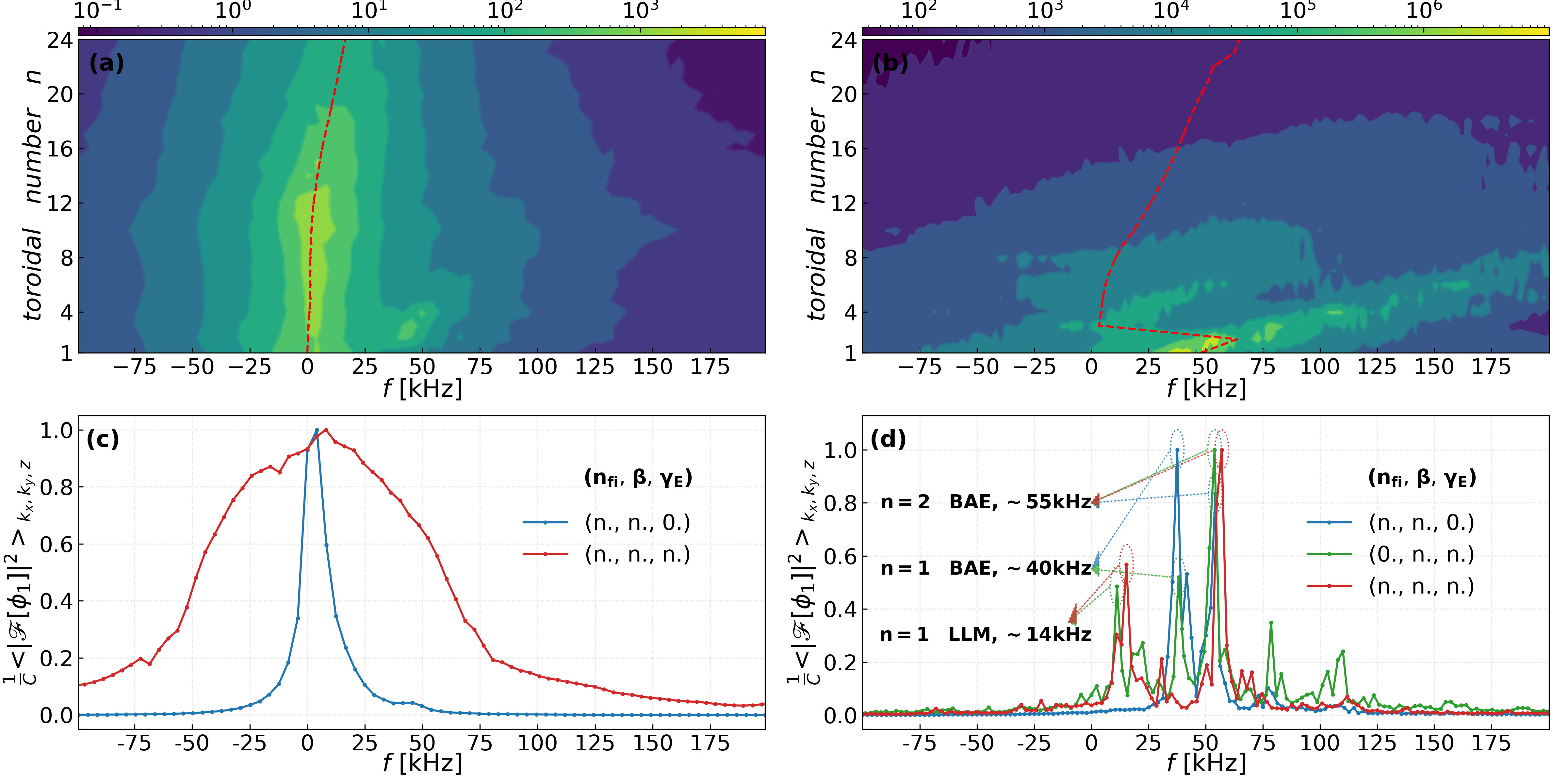}
    \caption{Frequency spectra of $\phi_1$ in nonlinear simulations. Logarithmic $ \langle |F[\phi_1] |^2 \rangle _{k_x,z} $ for the case with $(n_{fi}, \beta, \gamma_{E} )=(n., n.,0.)$ at (a) 510 ms and (b) 650 ms; red dotted line indicating the linear frequencies are aslo shown for the respective case in (a) and (b);normalized $\langle |F[\phi_1] |^2 \rangle _{k_x,k_y,z}$ at (c) 510 ms and (d) 650 ms for cases as labeled in the legend.}
    \label{nl_inst}
\end{figure*}

\begin{table}
\caption{\label{tab:parameters}Input parameters derived from HL-2A discharge \#22453 at two experimetal time 510 and 650 ms respectively. Parameters of electron are used for reference. For 510 ms, $n_e$=1.272$\times 10^{19} m^{-3}$, $T_e$=0.811 keV, and $B_t$ = 1.349 T, while for 650 ms, $n_e$=1.271$\times 10^{19} m^{-3}$, $T_e$=0.853 keV, and $B_t$ = 1.345 T. The aspect ratio R/a=4.87 and major radius R=1.68 m. In GENE code, electron $\beta_e = 8 \pi n_e T_e /B^2_t$ serves as a reference value and the total $\beta = (1+\sum_{j \neq e} n_jT_j/n_eT_e) \beta_e$. For the case without fast deuterium ion, its density $n_{fi}$ is set to zero and that of thermal deuterium ion, $n_{i}$, is set equal to $n_{e}$ to ensure quasi-neutrality. }
\begin{ruledtabular}
\begin{tabular}{lcccccc}
$t_{exp}$[ms] & $n_i/n_e$ & $n_{fi}/n_e$ & $T_i/T_e$ & $T_{fi}/T_e$ & $R/L_{n_e}$ & $R/L_{n_i}$\\
510 & 0.75 & 0.25 & 0.84 & 22.62 & 4.36 & 5.68 \\
650 & 0.72 & 0.28 & 1.29 & 22.22 & 6.83 & 9.13 \\
\hline
$t_{exp}$[ms] & $R/L_{n_{fi}}$ & $R/L_{T_{e}}$ & $R/L_{T_{i}}$ & $R/L_{T_{fi}}$ & $\beta_e[\%]$ & $\gamma_{E}$ \\
510 & 0.40 & 7.00 & 9.53 & 1.47 & 0.23 & 0.065 \\
650 & 0.80 & 7.22 & 25.61 & 1.67 & 0.24 & 0.187 \\
\end{tabular}
\end{ruledtabular}
\end{table}

\begin{figure*}
	\includegraphics[width=0.9\textwidth]{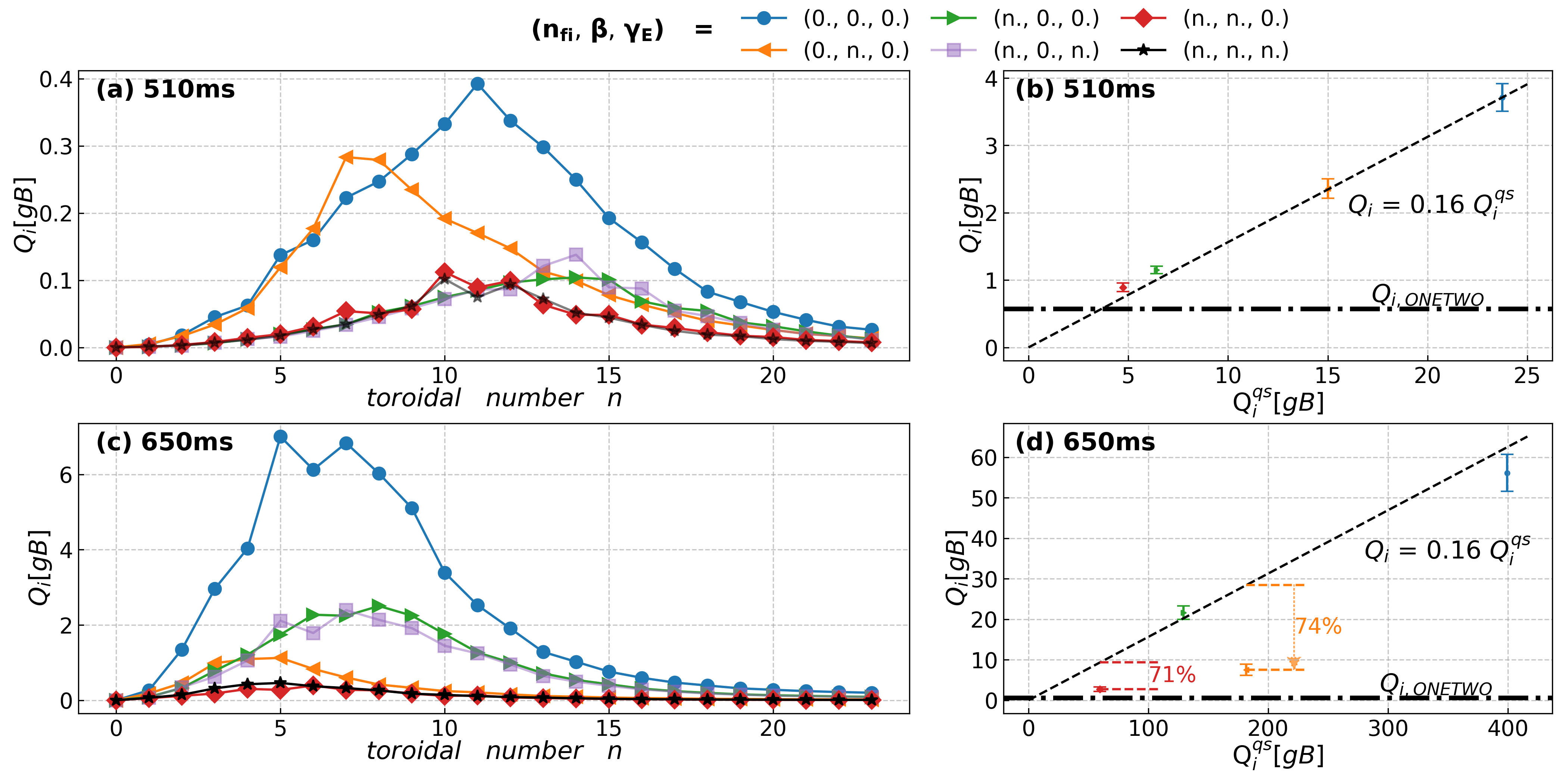}
	\caption{Spectra of ion heat fluxes for the cases at time 510 (a) and 650 ms (c) with $(n_{fi}, \beta, \gamma_{E} )$ as labeled above; the spectra at $n>24$ are omitted for they are low in values; total non-linear fluxes v.s. quasi-linear fluxes at time 510 (b) and 650 ms (d). The levels of ion power balance heat flux calculated by ONETWO, 0.58 and 0.54 gB, are also shown in (b) and (d) respectively.}
	\label{flux_sptrm}
\end{figure*}
\section{Instabilities}
The frequencies and growth rates of the most unstable modes in linear simulations are presented in Fig. \ref{linear_instabilities_new}. At both 510 and 650 ms, the spectra are dominated by the electrostatic (ES) ITG modes, which are characterized by frequencies in direction of ion diamagnetic drift and peaks at binormal wave number $k_y\rho_i \approx 0.3$ (or equally toroidal number $n \approx 12$). It can be observed, comparing the cases with and without $\beta$, that ITG is stablized by the well-known linear finite-$\beta$ effects \cite{weiland1992electromagnetic,hirose2000finite}. Furthermore, fast ions exert another damping effect on ITG. This damping effect arises from the dilution of main ion species, which act as the driven force of ITG, and thus reduces ITG growth rates by a factor scaling with the fast-ion concentration $n_{fi}/n_e$ \cite{tardini2007thermal, wilkie2018first}. After the ITB is well developed at 650ms, modes at toroidal number n=1 and n=2, with frequencies higher than those of ITG modes, are found destabilized without the contribution of the fast ions but rather as a consequence of the combined influence of steep $R/L_{T_{i}}$, low $\hat{s}$ and finite-$\beta$. It was shown \cite{Lin2022} previously that these EM modes have linear properties in good agreement with those of Beta-induced Alfvén Eigenmode (BAE) \cite{zonca1996kinetic}. As analyzed by linear simulations therein\cite{Lin2022}, these BAEs are mainly destabilized by the thermal ion temperature gradient with the critical value as 
\begin{eqnarray}
R/L_{T_i} \vert_{critical}=\frac{1}{q_0 \sqrt{7/4+T_e/T_i}} \frac{\omega_{tr}}{\omega_{*n_i}} \frac{R}{L_{n_i}},
\end{eqnarray}
where $\omega_{tr}=\sqrt{2T_i/m_i}/(q_0R)$ is the thermal ion transit frequency and $\omega_{*n_i}$ is the density part of the ion diamagnetic drift frequency $\omega_{*p_i}=k_y T_i (R/L_{n_i}+R/L_{T_i})/(e B_t R)$. In brief, the distinct property of these modes is that their frequencies scale with both the transit and diamagnetic drift frequency of thermal ion, $\omega \sim \omega_{tr} \sim \omega_{*p_i}$, nearly independent on the characteristic parameters of fast ion. Destabilized above a relatively low critical $\beta$, their mode structures in ballooning representation, unlike that of ITG modes which localized within small ballooning angle, have not only a large extension over the ballooning angle but also small scale variations with characteristic length of the order of $ \beta^{1/2}$.
\par
To gain an insight of the nonlinear characteristic of the instabilities, Fourier transforms are applied to fluctuating ES potential $\phi_1$ in the saturated phase of nonlinear simulations, and the results of several typical cases are averaged spatially and presented in Fig. \ref{nl_inst}. The zonal component of $\phi_1$, being the most prominent modes with zero frequency, are neglected in these spectra to highlight modes at $n \neq 0$. Two patterns of spectrum are generally observed comparing Fig. \ref{nl_inst}(a) and (b). For those cases where finite-$\beta$ and large $R/L_{T_{i}}$ are not jointly present, the nonlinear spectra are consistent with the linear results. As shown by the representative case in Fig. 3 (a), the peaks of the Fourier amplitudes coincide with the linear frequencies of most unstable modes, with bandwidths arising from nonlinear scattering of dominant modes or coexisting subdominant ones. For such cases, no modes are found prominent in frequency range different from those of ITG modes, except that the bandwidth, as can be seen in Fig. \ref{nl_inst}(c) is broadened with fintie $\gamma_E$. However, when finite-$\beta$ is considered in the presence of large $R/L_{T_i}$, high frequency modes appear apart from the ITGs. Those at n=1 and n=2 are the aforementioned BAEs with frequencies of around 40 and 55 kHz respectively, while those at higher $n$ have exponentially low amplitudes. It can be seen that, if the rotation frequency $f_{tor} \approx 6.7$ kHz is considered with $f=f_{lab}-nf_{tor}$, BAE at n=2 with $f \approx 55$ kHz corresponds to the mildly destabilized perturbation $f_{lab} \approx 70$ kHz in Fig. \ref{LLM_in_experiment}. When $\gamma_E$ is retained, as can be seen in Fig. \ref{nl_inst}(d) where the Fourier spectrum are averaged further over toroidal numbers, modes at n=1 appear with frequencies of $f \approx 14$ kHz, very close to the frequency of n=1 LLM in stationary frame. While finite $\gamma_E$ is essential for LLM to appear on one hand, fast ions act as a non-resonant energy source for its destabilization \cite{Zhang2014,Xie2022} on the other. As can be seen in Fig. \ref{nl_inst}(d), n=1 LLM could appear but would be dominated by n=1 BAE, if the contribution of fast ions were excluded. Most importantly, the presence of a significantly large $R/L_{T_i}$ is indispensable for the destabilization of LLM. \par
\begin{figure*}
	\includegraphics[width=0.95\textwidth]{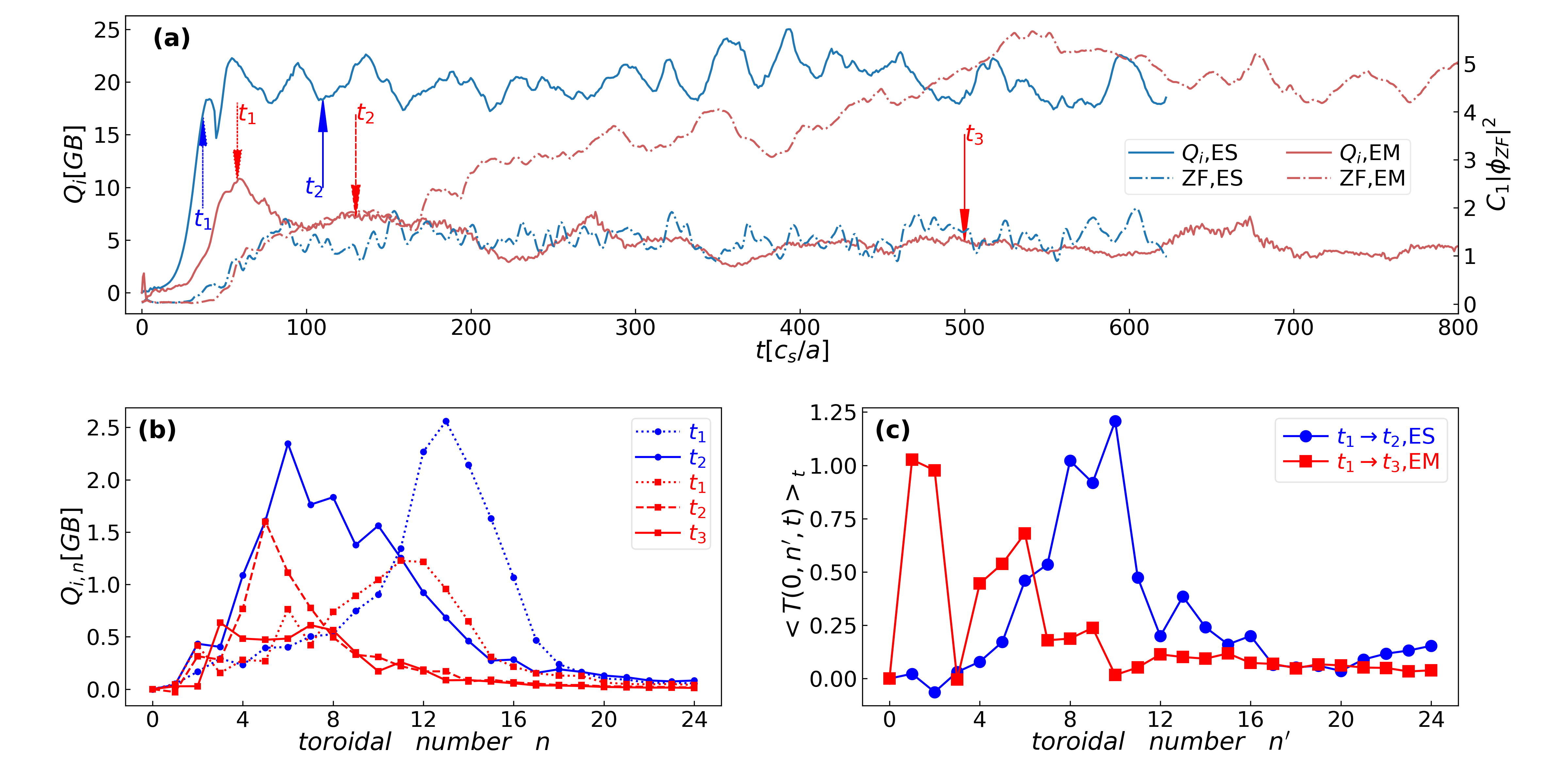}
	\caption{(a) Time trace of the total ion fluxes and ZF energy, labelled by $Q_i$ and $ZF$ respectively, for the cases at 650 ms with $(n_{fi}, \beta, \gamma_{E} )$ equal to $(n.,0.,n.)$ and $(n.,n.,n.)$, labelled by EM and ES respectively. (b) Flux spectrum at the normalized time marked in (a). (c) Time-averaged nonlinear term contribution from each $n^{\prime}$ to the growth of ZF energy for the EM and ES case during the labelled normalized time.}
	\label{zf_growth}
\end{figure*}
\section{Ion heat fluxes}
The toroidal spectrum of the flux-surface-averaged ion heat fluxes, computed from the saturated phase of nonlinear simulations in unit of gyroBohm (gB $\equiv n_e T_e^{5/2} m_i^{1/2} / (e^2 B_t^2 a^2)$), are presented in Fig. \ref{flux_sptrm}(a) and (c). For comparison, they are integrated over toroidal number $n$ and plotted in the right panels as a function of the quasi-linear ion heat fluxes calculated from the corresponding linear growth rates according to \cite{whelan2018nonlinear}. In addition, a model constant $\mathcal{C}$ is determined by the ratio of non-linear to quasi-linear flux level of the full case at 510 ms, and a dashed line indicating the prediction of quasi-linear model $Q_i=\mathcal{C}Q^{qs}_{i}$ is shown in Fig. \ref{flux_sptrm}(b) and (d). Quasi-linear theory contains no information of the exact nonlinear couplings between instabilities, typically the couplings with zonal components. Therefore, comparison between quasi-linear and non-linear fluxes disentangles the non-linear behaviours of the instabilities from the linear ones, and thus serves as an indicator of the nonlinear effects. In the same panels, the ion heat fluxes calculated by the transport code ONETWO \cite{pfeiffer1985giant} are shown by dash-dotted lines to indicate the power balance level. Also, the heat fluxes of fast ions are typically negligible compared to those of thermal ions, and therefore omitted in the following analysis. At 510 ms, it can be seen from Fig \ref{flux_sptrm}.(a) that the fluxes are significantly reduced when fast ions and/or finite-$\beta$ are included and that the effect of fast ions is more effective for the reduction. Comparing the full case and the case without any factor, the fluxes are found to be reduced by about 78\%, to which fast ion alone contribute about 90\% . Fast ions in our cases have large density but relatively low pressure gradient, destabilizing no extra mode as analyzed above. Their introduction in our simulations merely cause a dilution of the fraction of thermal ions, the latter being the main drive of ITG responsible for most of the fluxes. Such fast-ion dilution effect is closely pertinent to tokamak with low plasmas density and high NBI power, and is recognized as the key factor for the triggering of ITB at 510 ms. Another noteworthy point is that, as can be seen in Fig. \ref{flux_sptrm}(b), the non-linear fluxes are highly predictable by quasi-linear theory from the growth rates in the corresponding linear cases. This indicates that the mechanisms behind the flux reduction at 510 ms mainly lies in the linear stabilization effects of both fast-ion dilution and finite-$\beta$ as shown in Fig. \ref{linear_instabilities_new} (a).\par
The quasi-linear prediction are reliable also at 650 ms before finite-$\beta$ is included. In Fig. \ref{flux_sptrm}(c), ITG modes are fully destabilized due to steep $R/L_{T_i}$ and the fluxes driven by them, peaking around n=6 or $k_y \rho_i=0.15$, would reach 50 gB in total if all the stabilizing factors were discarded. The effect of fast-ion dilutions are not sufficient to suppress such fluxes. Instead, it is only when finite-$\beta$ is retained that the fluxes are drastically reduced. More importantly, the flux reduction caused by finite-$\beta$ in nonlinear simulations largely overtakes what is predicted by quasi-linear model. It is seen in Fig. 4(d) that the effect of finite-$\beta$ alone can reduce the total 50 gB fluxes to 7.5 gB, deviating from the quasi-linear prediction by about 74\%. When all the other factors are considered along with finite-$\beta$, the total fluxes are eventually reduced to around 2.7 gB. Non-linear finite-$\beta$ effect is thus identified as the dominant stabilizing effects during the sustainment phase of ITB. The underlying mechanism was investigated previously in \cite{whelan2018nonlinear} and attributed to an energy transferring enhanced by finite-$\beta$ between the flux-driven ITG modes and zonal flows. As will be shown below, a link between the flux reduction and the prominent growth of zonal flows is indeed identified with finite-$\beta$, but the energy required for such growth are mainly tapped from n=1 EM modes instead of the ITG ones. \par
The inclusions of both fast ion and finite-$\beta$ have made a significant contribution to drawing the simulated heat fluxes near the power balance fluxes, but finite difference still exist between them at both 510 and 650 ms. Such differences may arise from the inevitable errors in measurement and parameters evaluations, but they could do little harm to our conclusions which rely mainly on the relative change of fluxes rather than on their absolute values under different simulation conditions. By sharp contrast to the beneficial role played by fast ion and finite-$\beta$, the effects of $E \times B$ shearing on the fluxes are barely visible at both 510 and 650 ms. It is found in Fig. \ref{flux_sptrm} that the retaining of finite $\gamma_{E}$ slightly increased the fluxes, but such change is within the error bar. The ineffectiveness of $\gamma_{E}$ was reported also in other tokamaks \cite{citrin2014electromagnetic, pan2017investigation} and can be simply explained by its low value, which in our case is only about one third of the growth rates of dominating ITG modes (if compared in the same unit $c_s/a$). Therefore, it is concluded that the $E \times B$ shearing has not direct impact on the ITB formation other than changing the characteristics of the n=1 EM modes, which, in spite of their dominant amplitudes in frequency spectrum, drive much less fluxes than their ITG companions.\par
\section{Flux reduction and zonal flows}
The aforementioned discrepancies between non-linear and quasi-linear fluxes with finite-$\beta$ are related to the effect of the zonal flows (ZF). To confirm this, the full time traces of total ion fluxes are shown in Fig. \ref{zf_growth}(a) for the cases with and without finite-$\beta$ at 650 ms, labelled by EM and ES respectively, and in the same plot the ZF energies for the respective case are also displayed. Here, the field energy of each $n$ is defined concerning only the ES part as $E_n \equiv \sum_{k_x} \int Jdz C_1 |\phi_1|^2$, where $J$ is the Jacobian. A positive-defined real constant $C_1(k_x,k_y,z)$ is included so that $E_n$ corresponds to the field part of the free energy \cite{banon2011free}, and could be substituted with other positive-defined real constant such as $k_{\perp}^{2}$. From Fig. \ref{zf_growth}(a) , it’s observed regardless of whether finite-$\beta$ is retained or not, that the ion heat fluxes at first develop linearly to form the $\gamma_n$-dependent shape during the initial phase, the time window before $t_1$ when the amplitudes of ZFs are low, and that the peaks of the spectra begin to shift toward lower toroidal number as the ZFs continue to develop in the transient phase from $t_1$ to $t_2$. The corresponding flux spectra are shown in Fig. \ref{zf_growth} (b). At this phase, the heat flux spectra tend to evolve into the $\gamma_n/k_{\perp}^{2}$-dependent \cite{Jenko2005heat,bourdelle2007a} quasi-linear shape with overall values predictable from the dashed line in Fig \ref{flux_sptrm}(d). For the ES case, the heat fluxes begin to saturate at the quasi-linear level. For the EM case, however, it is observed that the heat fluxes, instead of becoming saturated, continue to abate slowly as the ZF energy in such case is experiencing a persistent growth, whose cause is reported in the following. The time evolution of $E_n$ naturally depends on the linear contributions and nonlinear ones, but only the latter contribute to the net growth of ZF energy $E_0$.  Taking the time derivative of $E_n$ and substituting $\phi_1$ with the modified distribution function $g_1$ through the field equation, the nonlinear contribution to $dE_n/dt$ is expressed as (see the Appendix)\\ 
\begin{eqnarray}
\begin{aligned} \label{eq:nl_ev} &\frac{dE_n}{dt} \vert_{NL}
\equiv \sum_{n^{\prime}} T(n, n^{\prime},t)\\
&=Re\sum_{k_y^{\prime},k_x,k_x^{\prime}} (k_x^{\prime}k_{y}- k_y^{\prime} k_x) \int Jdz M \bar{\phi}_1^* |_{k} \chi_{1j} |_{k^{\prime}} g_{1j} |_{k-k^{\prime}}, 
\end{aligned}
\end{eqnarray}
where $M= \sum_j \pi n_j q_j  \int dv_{\parallel} d \mu B_0 $ is a moment operator and $\chi_1=\bar{\phi}_1 - v_{th,j} v_{\parallel} \bar{A} _{1,\parallel} + T_{j} \mu / q_j \bar{B}_{1\parallel}$ the gyro-averaged effective potential. $T(n,n^{\prime},t)$ is calculated focusing on the coupling between ZF ($n=0$) and all the other $n^{\prime}$, and the results shown in Fig. \ref{zf_growth}(c) have been averaged over the time window when there is a net growth of ZF. It is thus seen that ZF has mainly drained energies from $n=8 \sim 12$ ITG components when the low-n EM modes are artificially suppressed by neglecting finite-$\beta$ effect. Instead, when finite-$\beta$ is retained and the low-n EM modes are destabilized by the large $R/L_{T_i}$, ZF receives a significant positive portion of energy from these low-n modes, mostly from $n=1$ LLM, and develops to a much larger amplitude than that in the case without finite-$\beta$. Such favorable energy transfer is an evidence of the self-regulatory system where an EM mode, serving as a catalyst, transfers the free energy it obtained from the ITB-generated large $R/_{T_i}$ to the ZF which helps mitigate the heat fluxes and in turn sustain the ITB. Consequently, a self-organized mechanism is proposed which is characterized by an energy transfer that is facilitated by the saturation of the low-n EM modes, in this case the LLM, and that results in the increase of ZF activities and reduction of heat fluxes.\par
\section{Discussion and conclusions}
The ITB characteristics in HL-2A have been analyzed by performing non-linear gyrokinetic simulations. The emphasis of our study have been placed on the effects of fast ions, finite-$\beta$ and $E \times B$ shear. It is found that the complete ITB formation process can be conceptually divided into two stages where distinct mechanisms dominate. Widely effective as it is, the $E \times B$ shear stabilization in our cases is not found to play a remarkable role on any of these stages, mainly because the shearing rate $\gamma_E$ is ralatively low compared to the ITG growth rates. On the first stage, the plasmas instabilities are dominated by ITG modes which are subjected to the stabilizing effects of both finite-$\beta$ and fast ions. It is found that the triggering of ITB is mainly caused by the stabilization of ITG under the effect of fast-ion dilution which basically depend on linear physics. Once the ITB is fully developed, the sustaining of the ITB is determined by the reduction of heat turbulent transport by large scale zonal flows. On this second stage, the steep ITB-generated pressure gradient, combined with the effect of finite-$\beta$, is able to bring about an abundant varieties of large scale EM modes, in our case the LLM. Instead of driving significant fluxes, LLM acts as a catalyst that transfers the ITB free energy obtained during the triggering process to the zonal flows, which in turn mitigate the flux and sustain the ITB ultimately. The full ITB formation is therefore characterized as a self-regulatory multi-scale physics system leading to a self-sustained ITB. Although these conclusions are obtained from simulations which employ several simplifying model, such as the local assumption and Maxwellian fast ions, they provide an initial picture of the process of ITB formations. To validate our conclusions in further, the global gyrokinetic simulations that is much more demanding computationally may be necessary to rigorously account for the effect of  large scale flow shear and to completely accommodate all the modes involved. Nevertheless, the mechanism proposed in this paper could be important, e.g. if LLM can be induced, e.g. by tailoring the q-profile, to future tokamak devices like ITER with low $E \times B$ shearing, which is not found to play a major role here on any stage of ITB formation. \par
\section{Acknowledgement}
The authors are very grateful to Mr. Chen Qian, Mr. Zhang Xing, Mr. Fang Kairui, Dr. Hao Guangzhou, Dr. Yu Deliang and the HL-2A experiment team for providing and processing experimental data. This work was supported by the National Natural Science Foundation of China with grant Nos. 12275071 and U1967206 and also partially by National Key R\&D Program of China under Grant Nos. 2017YFE0301200 and 2017YFE0301201.\par
\section{Appendix}
The derivations of Eq. \ref{eq:nl_ev} are reported in the following. In flux-tube version of GENE, the normalized gyrokinetic Vlasov equation for the modified distribution function $g_{1j}$ of species $j$ can be written as
\begin{eqnarray}
\frac{dg_{1j}}{dt}&=&L_G \chi_{1j} + L_C ( g_{1j} + q_j \chi_{1j} \frac{F_{0j}}{T_j} ) \nonumber\\
&&+ L_{\parallel}( g_{1j} + q_j \chi_{1j} \frac{F_{0j}}{T_j} ) - \{ \chi_{1j}, g_{1j}\}_{x,y},
\label{eq:gyv}
\end{eqnarray}
where $L_G$ is the gradient prefactor,
\begin{eqnarray}
L_G=((-\frac{3}{2} + v^2_{\parallel} + \mu B_0) \frac{R}{L_{T_j}} + \frac{R}{L_{n_j}})F_{0j} i k_y,
\label{eq:lg} 
\end{eqnarray}
$L_C$ the curvature prefactor,
\begin{eqnarray}
&L_C&=\frac{T_{j}( 2v^2_{\parallel} + \mu B_0)}{q_jB_0} K_x i k_x
\nonumber\\
&&+(-\frac{T_{j}( 2v^2_{\parallel} + \mu B_0)}{q_jB_0} K_y +
\beta \frac{T_j v^2_{\parallel} p_0}{q_j B_0^2} \frac{R}{L_{p_0}} )ik_y,
\label{eq:lc} 
\end{eqnarray}
and $L_{\parallel}$ the parallel-dynamic operator,
\begin{eqnarray}
L_{\parallel}=v_{th,j} \frac{F_{0j}}{2} \frac{1}{JB} \{ \frac{1}{F_{0j}},\quad \}_{v_{\parallel},z}.
\label{eq:lpar} 
\end{eqnarray}
Here, the Poisson bracket of two arbitrary function $f$ and $g$ over the variables $u$ and $v$ is defined as 
\begin{eqnarray}
\{f, g \}_{u,v}=\frac{\partial f}{\partial u}\frac{\partial g}{\partial v} - \frac{\partial f}{\partial v}\frac{\partial g}{\partial u}.
\label{eq:psb}
\end{eqnarray}
When the nonlinear term in Eq. \ref{eq:gyv} (the last term in the right hand side) is evaluated in Fourier space at $(k_x, k_y, z)$, the multiplications in the Poisson bracket are transformed into convolutions, i.e.
\begin{eqnarray}
\begin{aligned}
\quad & -\{ \chi_{1j}, g_{1j}\}_{x,y} \vert_{k}\\
& \quad = \sum_{k_x^{\prime},k_y^{\prime}}
(k_x^{\prime}k_y-k_xk_y^{\prime}) \chi_{1j} \vert_{k^{\prime}} g_{1j}\vert_{k-k^{\prime}}.
\label{eq:nlt}
\end{aligned}
\end{eqnarray}
Full magnetic fluctuations are considered in $\chi_1=\bar{\phi}_1 - v_{th,j} v_{\parallel} \bar{A} _{1,\parallel} + T_{j} \mu / q_j \bar{B}_{1\parallel}$, the gyro-averaged effective potential, where the bar over quantities indicates gyro-average. Note that in local limit the gyro-average of ES potential $\phi_1$ is simply $\bar{\phi}_1=J_0(k_{\perp} \rho_j)\phi_1$, where the $J_n$ is the Bessel function of $n^{th}$ order. The symbol $\vert_k$ indicates the quatity before it is evaluated at $(k_x,k_y)$. In the curvature term, the $K_x$ and $K_y$ are the curvature factor in radial and binormal direction respectively. Their definition, as well as those of other quantities, can be found in, e.g. ref. \cite{merz2008gyrokinetic} and \cite{told2012gyrokinetic}. The field equation of the ES potential $\phi_1$ is coupled with that of the parallel fluctuating magnetic field $B_{1\parallel}$ when finite-$\beta$ is considered. The coupled field equations are
\begin{eqnarray}
C_1 \phi_1 + C_2 B_{1\parallel} & = & M J_0(k_{\perp} \rho_j) g_{1j},
\\ \label{eq:field1}
C_2 \phi_1 + C_3 B_{1\parallel} & = & M \frac{2J_1(k_{\perp} \rho_j)}{k_{\perp} \rho_j)} \frac{T_{j} \mu}{q_j} g_{1j}, 
\label{eq:field2}
\end{eqnarray}
from which we obtain 
\begin{eqnarray}
\phi_1 &=&\frac{C_3}{C_1C_3 - C_2^2} M J_0(k_{\perp} \rho_j) g_{1j} 
\nonumber\\
&& - \frac{C_2}{C_1C_3 - C_2^2} M \frac{2J_1(k_{\perp} \rho_j)}{k_{\perp} \rho_j)} 
\frac{T_{j} \mu}{q_j} g_{1j}.
\label{eq:field_phi}
\end{eqnarray}
The moment operator $M$ is defined as
\begin{eqnarray}
M = \sum_j \pi n_j q_j  \int dv_{\parallel} d \mu B_0,
\end{eqnarray}
and the definitions of the coefficients $C_1$, $C_2$ and $C_3$ (which are real and only depend on $k_x$, $k_y$ and $z$) can be found in Page 33 of ref. \cite{merz2008gyrokinetic}. By Eq. \ref{eq:field_phi}, the total derivative of the ES energy at $k_y$ can be expressed as\\
\begin{eqnarray}
\begin{aligned}
\frac{d E_n}{dt} 
&= \sum_{k_x} \int Jdz C_1 (\frac{ \partial \phi_1^{*} }{\partial t} \phi_1 + \phi_1^{*} \frac{ \partial \phi_1 }{\partial t})
 \\
 &= \; Re \sum_{k_x} \int Jdz C_1 \phi_1^{*} \frac{ \partial \phi_1}{\partial t}
\\
&= \; Re (\sum_{k_x} \int Jdz \frac{C_1}{C_1C_3 - C_2^2} \phi_1^{*}
\\
& \times M( C_3 J_0(k_{\perp} \rho_j) - C_2 \frac{2J_1(k_{\perp} \rho_j)}{k_{\perp} \rho_j} \frac{T_{j} \mu}{q_j} )\frac{ \partial g_{1j}}{\partial t} .
\end{aligned}
\label{eq:tot_dEdt}
\end{eqnarray}
As the nonlinear contribution alone is of our concern, we can obtain, by substituting the $\partial g_{1j}/\partial t$ in Eq. \ref{eq:tot_dEdt} with only the nonlinear term Eq. \ref{eq:nlt}, 
\begin{eqnarray}
\begin{aligned}
\frac{dE_n}{dt}\vert_{NL}=&Re\sum_{k_y^{\prime},k_x,k_x^{\prime}}(k_x^{\prime}k_{y}-k_y^{\prime} k_x)\int Jdz M \phi_1^* \vert_{k}
\\
&\times (\frac{C_1 C_3}{C_1C_3 - C_2^2} J_0(k_{\perp} \rho_j) 
\\& \quad - \frac{C_1C_2}{C_1C_3 - C_2^2}\frac{2J_1(k_{\perp} \rho_j)}{ k_{\perp} \rho_j}) 
\\
& \times \chi_{1j} \vert_{k^{\prime}} g_{1j}\vert_{k-k^{\prime}},
\end{aligned}
\label{eq:nl_dEdt_1}
\end{eqnarray}
where the commutativity between the moment operator $ M $ and any spatial quantities has been used. As the coefficient $ C_3 $ is inversely proportional to $\beta$ which is closed to zero for our case, the second term in Eq. \ref{eq:nl_dEdt_1} makes negligible contribution to the total value. Therefore, taking the limit $C_3 \rightarrow \infty$, Eq. \ref{eq:nl_ev} can be obtained from Eq. \ref{eq:nl_dEdt_1}. But note that Eq. \ref{eq:nl_dEdt_1} instead of \ref{eq:nl_ev} is actually used to produce Fig. \ref{zf_growth} (c) for the sake of completeness.

\bibliography{HL2A_ITB}

\end{document}